# Spatiotemporally unified air-sea interaction in tropical oceans

Yaokun Li[1]*

[1]College of Global Change and Earth System Science, Beijing Normal University, Beijing, 100875, China.

*Corresponding author: Yaokun Li (liyaokun@bnu.edu.cn)

**Abstract**

The spatiotemporal variation in tropical air-sea interaction is investigated by applying a simple model that considers the fundamental dynamics in tropical oceans. The model decomposes sea surface temperature anomaly (SSTA) variation into a series of spatial modes that oscillates with their natural frequencies. The results suggest that the first mode associates with the dipole-like SSTA variation between the western and the eastern coast, such as EP El Niño, the Atlantic Niño, and IOD; whereas the second mode associates with the tripole-like SSTA pattern among the central and eastern, western coast, such as CP El Niño and minor SSTA variations in the tropical Atlantic and Indian Ocean. Each mode oscillates with its natural frequency that depends on the strength of air-sea coupling and the basin size. The model provides a systematic framework for the comprehensive understanding of the complex air-sea interaction in tropical oceans.

# 1 Introduction

Large scale air-sea interaction plays a crucial role in climate change and variability on a broad range of time scales (Neelin et al., 1992). In the tropical Pacific, the El Niño/Southern Oscillation (ENSO) cycle, accompanied with alternating warm El Niño and cold La Niña events, is the most prominent year-to-year climate variation on Earth (McPhaden et al., 2006). In the tropical Atlantic, the Atlantic Niño, featured by the appearance of warm sea surface temperature anomalies (SSTAs) in the eastern equatorial Atlantic in northern summer, is analogous to but weaker than El Niño (Zebiak, 1993). In the tropical Indian Ocean, the Indian Ocean Dipole (IOD), covered by the basin-wide uniform SSTA variation, describes the difference in SSTA between the tropical western and south-eastern Indian Ocean and is also analogous to but weaker than El Niño (Saji et al., 1999).

The fundamental mechanisms of the tropical air-sea interaction family are the same. Each develops through Bjerknes positive feedback (Bjerknes, 1969) that suggest the warm SSTA in the eastern basin can weaken the trade wind and hence the upwelling, and can eventually amplify the warm SSTA in the eastern basin (Zebiak, 1993; Iizuka et al., 2000; Gualdi et al., 2003; Keenlyside & Latif, 2007; Ng et al., 2014). In the tropical Pacific, the Bjerknes positive feedback is balanced by corresponding negative feedback mechanism to form the complete ENSO cycle. Negative feedbacks, such as the delayed oscillator (Suarez & Schopf, 1988; Battisti & Hirst, 1989) that emphasizes the western boundary's reflection of Rossby and Kelvin waves and the recharge oscillator (RO) (Jin, 1997) that emphasizes the recharge-discharge of the equatorial heat content (Cane & Zebiak, 1985; Wyrtki, 1985), had been proposed to explain the phase transition mechanism of ENSO. In the tropical Atlantic and Indian Ocean, these classic negative feedbacks can also be applied although their influences may be modulated by other mechanisms (Rao et al., 2002; Jansen et al., 2009; McPhaden & Nagura, 2014; Prigent et al., 2020).

Each member of the tropical air-sea interaction family has complex spatiotemporal variations (Weller et al., 2014; Timmermann et al., 2018; Vallès Casanova et al., 2020). The differences in the zonal structure and strength of air-sea coupling and mean ocean stratification may explain the specific features (Zebiak, 1993). Complex interactions among these brothers may also contribute to this complexity (Kajtar et al., 2017; Cai et al., 2019). Besides, basin size may also matter. For example, since the basin size of the equatorial Atlantic is around one half that of the Pacific, the Atlantic Niño has a considerably shorter period of quasi-biennial timescale than ENSO of a dominant quasi-quadrennial timescale (Latif & Groetzner, 2000). For another example, since the east-west contrast of the thermocline depth depends on the integrated wind stress over the basin, the intensity of the Atlantic cold tongue is significantly weaker in the eastern Atlantic, also about one half of that in the eastern Pacific (Jin, 1996).

Previous advances call for a systematic understanding of the analogous nature but dramatic spatiotemporal discrepancy in the complex tropical air-sea interaction phenomenon. Taking the tropical Pacific for an example, the RO framework has achieved great success in understanding physical processes that determines ENSO's temporal variations (Jin et al., 2020). However, it cannot depict the diversity in the spatial patterns, known as eastern Pacific (EP) El Niño and central Pacific (CP) El Niño, which vary with different significant oscillation periods (Xie & Jin, 2018). Therefore, it is a main challenging issue to combine ENSO's spatiotemporal variations in a simple concept model (Jin, 2022). In this investigation, a simple but physically based model is built in the tropical oceans to complete this challenging task.

**2 The theory**

SSTA in the tropical ocean is governed by the advection terms, the vertical advection terms and net surface heat flux. To highlight the fundamental feedback process between SSTA and the thermocline depth anomaly (TDA), SSTA equation can be simplified to

$$\frac{\partial T}{\partial t} + K_h h = 0, \qquad (1)$$

where $T$ and $h$ denote SSTA and TDA, respectively, the parameter $K_h$ represents the linearized relationship between SSTA and TDA and varies both zonally and meridionally (Wakata & Sarachik, 1991). Despite that, it can be specified to a constant value at the first glance. Of course, its value should be different in different basins. The more $K_h$ increases, the more tilt the thermocline will become and hence the stronger the air-sea coupling will become (Wakata & Sarachik, 1991). Eq. (1) has been widely applied in theoretical studies (Neelin & Jin, 1993; Jin & Neelin, 1993; Kang & An, 1998; Kang et al., 2001). Note that the horizontal advection terms have been ignored to maximum simplify the equation. It is reasonable since they are small compared to the upwelling (Battisti & Hirst, 1989).

TDA is governed by the zonal advection by oceanic dynamics

$$\frac{\partial h}{\partial t} + H \frac{\partial u}{\partial x} = 0, \qquad (2)$$

where $H$ is the mean depth of the mixed layer, $u$ is the zonal oceanic current anomaly. Eq. (2) introduces the oceanic motion, an additional variable, to eliminate which the zonal momentum equation should be included. However, this will further complicate the system so that the analytic solution is hard to derive. Considering that the motions in the mixed layer is strongly driven by the wind stress, a simple relation can link the oceanic current anomaly with the wind stress anomaly, that is

$$u = b_1 \tau, \qquad (3)$$

where $b_1$ is the coupling coefficient and $\tau$ is the zonal wind stress anomaly.

Further considering that the zonal wind stress anomaly is determined by the zonal gradient in SSTA, a simple relation can also be proposed to link the zonal wind stress anomaly with the zonal gradient in SSTA, that is

$$\tau = b_2 \frac{\partial T}{\partial x}, \qquad (4)$$

where $b_2$ is the coupling coefficient. Eqs. (3) and (4) build the physical relationship between the oceanic current anomaly and the zonal gradient in SSTA. Applying this relation, a system that considers the fundamental dynamics in SSTA and TDA is enclosed. Note that the system can be regarded as a simplified version of the coupled system (Cane & Zebiak, 1985; Zebiak & Cane, 1987; Neelin & Jin, 1993; Jin & Neelin, 1993). Eliminating $h$, it is easy to derive a wave equation for SSTA

$$\frac{\partial^2 T}{\partial t^2} = c^2 \frac{\partial^2 T}{\partial x^2}, \qquad (5)$$

where $c = \sqrt{K_h H b}$ is the propagation speed of the air-sea coupled wave and $b = b_1 b_2$. According to the previous study (Hirst, 1986; Kang & An, 1998), $K_h$ has an order of $1.0 \times 10^{-8} \text{ K m}^{-1} \text{ s}^{-1}$. Observational-based analysis suggests that the Bjerknes feedback is up to 50% weaker in the Atlantic and in the Indian Ocean (Keenlyside & Latif, 2007; Jansen et al., 2009). Therefore, it is reasonable to hypothesize that the value of $K_h$ varies with the basin size to modulate the strength of the air-sea coupling. This will be discussed later. The depth of the mixed layer $H$ is set to be a constant 50m in all three tropical oceans although there exists discrepancy in each ocean. However, this discrepancy is not significant and can be included in the air-sea coupling strength. The parameter $b$ represents the relation between the zonal oceanic current anomaly and the gradient in SSTA. The typical gradient value in SSTA has an order of $10^{-6} \text{ K m}^{-1}$ and the zonal oceanic current anomaly has an order of $0.1 \text{ m s}^{-1}$. Therefore, the parameter has an order of $1 \times 10^5 \text{ m}^2 \text{ s}^{-1} \text{ K}^{-1}$. The warmer tropical SSTA is, the stronger wind anomaly it can drive and hence stronger oceanic current anomaly in the mixed layer. Therefore, this parameter does not depend on the basin size. With above values of parameters, the propagation speed of the air-sea coupled wave is $0.22 \text{ m s}^{-1}$.

# 3 The solution

## 3.1 Nondimensionalization

Let's first discuss the effect of the basin size by nondimensionalizing Eq. (5) before analytically solving it. Assuming the characteristic time scale is $\tau$ while the basin width is $L$, the dimensionless form of Eq. (5) is

$$\frac{\partial^2 \tilde{T}}{\partial \tilde{t}^2} = \beta^2 \frac{\partial^2 \tilde{T}}{\partial \tilde{x}^2}, \tag{6}$$

where variables with a superscript $\sim$ means corresponding dimensionless variables and $\beta = \frac{c\tau}{L}$ is a dimensionless parameter. It may be regarded as the ratio between the propagation distance ($c\tau$) of the air-sea coupled wave in the targeted time scale ($\tau$) and the basin width ($L$). For example, the interannual oscillations in tropical oceans are the most significant, which requires a proper basin width to balance Eq. (6). If the basin size is far less than the characteristic distance, $\beta$ will be far larger than unit so that the second derivative with respect to time can be ignored. Additional terms, such as the net surface heat flux, should be introduced to balance the left term. It implies that SSTA will only follow the external forcing with no internal oscillations, namely, no air-sea interaction. If the basin size is far larger than the characteristic distance, $\beta$ will be far less then unit so that its associated spatial variation term can be ignored. Similarly, additional terms also should be introduced to balance the equation. In this case, SSTA will only oscillate with the introduced terms but losing its natural spatial variation. Therefore, the dimensionless parameter $\beta$ may be thought as the scaler of the air-sea interaction in the tropical oceans. A proper value of around unit links the temporal oscillations with the spatial variations, corresponding to the significant air-sea interaction.

## 3.2 The cosine series expansion

The coupled system Eq. (5) can be solved by specifying proper initial values and free boundary conditions

$$T(x,0) \equiv f(x), \tag{7}$$

$$\left.\frac{\partial T}{\partial x}\right|_{x=0} = \left.\frac{\partial T}{\partial x}\right|_{x=L} = 0, \tag{8}$$

where $f(x)$ is a known function and $0 \leq x \leq L$ is the range of the basin. It is obvious that the cosine expansion in spatial dimension will naturally satisfy the boundary condition. Therefore, the solution is set to be

$$T = \sum_{n=0}^{\infty} T_n(t) \cos \lambda_n x, \qquad (9)$$

where $\lambda_n = \dfrac{n\pi}{L}$, $n = 0, 1, 2, 3, \cdots$. Correspondingly, the initial value become

$f(x) = \sum_{n=0}^{\infty} f_n \cos \lambda_n x$ where $f_n$ are expansion coefficient. Substituting Eq. (9) into Eq. (5) can derive

$$\frac{d^2 T_n}{dt^2} + \omega_n^2 T_n = 0, \qquad (10)$$

where $\omega_n = c \lambda_n$ is the natural frequency. It is easy to solve above equations and the final solution becomes

$$T = \sum_{n=0}^{\infty} \left( f_n \cos \omega_n t - \frac{K_h}{\omega_n} g_n \sin \omega_n t \right) \cos \lambda_n x, \qquad (11)$$

where $g_n$ is the expansion coefficient for initial TDA, namely,

$$h(x, 0) \equiv g(x) = \sum_{n=0}^{\infty} g_n \cos \lambda_n x.$$

**4 Results**

The solution (11) manifests that variations in SSTA have been decomposed into the summation of a series spatiotemporal modes ($n = 1, 2, 3, \cdots$) and a constant mode ($n = 0$). The first mode ($n = 1$), a standard cosine curve with a period of $2\pi$ in the ocean basin range $[0, L]$, features the SSTA contrast between the eastern and western basin (the blue curve in Fig. 1a). It is interesting to note that the dipole spatial pattern exists in all the three tropical oceans (Deser et al., 2010). For example, the empirical orthogonal function (EOF) analysis for SSTAs in the simple ocean data assimilation (SODA) version 3.4.2 (Carton et al., 2018) suggests that the major leading EOF mode is featured by a dipole pattern, e.g., the EP El Niño events in the tropical Pacific shown by the first leading EOF mode (the red curve in Fig. 1a), the Atlantic Niño in the tropical Atlantic shown by the second leading EOF mode (the yellow curve in Fig. 1a) and IOD in the tropical Indian Ocean shown by the second leading EOF mode (the purple curve in Fig. 1a). It is worthy to point out that these dipole patterns are quite consistent with the first mode (Fig. 1a). The second mode ($n = 2$), a standard cosine curve with a period of $\pi$ in the range $[0, L]$, highlights SSTA contrast between central and eastern, western basin (the blue curve in Fig. 1b). Also, the tri-pole spatial pattern exists in the tropical Pacific and Atlantic (Deser et al., 2010), e.g., the CP El Niño events in the tropical Pacific shown by the second leading EOF mode (the red curve in Fig. 1b), and the meridional mode characterized by cross-equator SSTA gradients in the tropical Atlantic shown by the third leading EOF mode (the yellow curve in Fig. 1b). In the tropical Indian Ocean, there also exists a tri-pole SSTA variation shown by the third EOF mode although its contribution is quite small (the purple curve in Fig. 1b). It is interesting to note that these tri-pole spatial patterns are quite consistent with the second eigenmodes. The constant mode, a constant value of unit, features consistent variation throughout the basin. This mode associates with the first leading EOF mode in the tropical Atlantic and Indian Ocean, manifesting its important role in determining the variations in SSTA.

An astonishing coincidence between the principle component (PC) time series and the cosine series expansion coefficient $T_n(t)$ for the SODA SSTA data as defined in Eq. (9) can also be seen in Fig. 2. In the tropical Pacific, the first and second PC time series associates with the first and second cosine series expansion coefficients; while in the tropical Atlantic and Indian Ocean, the second and third PC time series associate with the first and second series expansion coefficients.

The first mode oscillates with the fundamental frequency ($\omega_1 = c\frac{\pi}{L}$) and the other natural frequencies are the integral multiples of the fundamental frequency. Specific to the tropical Pacific, the fundamental period corresponding to the fundamental frequency is around 4 years and the second natural period is around 2 years. These two natural periods are close to the significant QQ and QB modes for the EP and CP El Niño events (Bejarano & Jin, 2008; Ren & Jin, 2013). Specific to the tropical Atlantic and Indian Ocean, the fundamental period is around 2 years, also very close to the observations (Deser et al., 2010; Zhang & Han, 2021). Therefore, the model provides a systematic and comprehensive physical view for understanding the tropical air-sea interaction and can be verified in the SODA data. Specific to the tropical Pacific, the first two modes are minimum required to feature its spatiotemporal variation. Specific to the tropical Atlantic and Indian ocean, the first mode and the constant mode are minimum required to depict their spatiotemporal variations.

To explicitly discuss the dependence of the fundamental period on the basin size. It is expressed as a function of the basin width

$$P_1(L) = \frac{2\pi}{\omega_1} = \frac{2L}{c(L)}. \tag{12}$$

It is obvious that the fundamental period is proportional to the basin width but inverse proportional to the air-sea coupled wave speed. According to above analysis, the air-sea interaction will disappear when the basin size is too small to breed its natural oscillation and will not increase infinitely when the basin size is too large. Therefore, an approximation expression is proposed to quantify this relation

$$K_h = \frac{1}{2} K_{h0} \left[ 1 + \tanh \sigma \left( \frac{L}{L_0} - 1 \right) \right], \quad (13)$$

where $K_{h0}$ corresponds to the strongest air-sea coupling strength, $L_0$ is a critical basin width, $\sigma$ is a parameter that controls the sharpness. If the basin width is much wider than the critical width, the parameter $K_h$ will tend to $K_{h0}$, realizing the strongest air-sea coupling strength. If the basin width is much narrower than the critical width, the parameter $K_h$ will tend to be zero, meaning no local air-sea coupling. This hypothesized relation may also be supported by the investigation of the Tertiary tropical climate variability in a prototype model (An et al., 2012), in which the opening Panama Seaway has no significant influence on the tropical Pacific due to the wide Pacific width, whereas, closed seaways lead to no interannual fluctuation in the Atlantic due to the narrow Atlantic basin. Therefore, $K_{h0}$ may be set to equal to the value in the Pacific. Furtherly, $L_0$ may be set to equal the Atlantic basin width $L_A$ so that $K_h$ will be the half of the strongest air-sea coupling strength, consistent with the previous analysis. The fundamental period is basically proportional to the basin width although $K_h$ increases from $\frac{1}{2} K_{h0}$ at $L_A$ to close $K_{h0}$ at $L_P$, the basin width of the tropical Pacific (Fig. 3a). Besides, the fundamental period will be modulated by sharp $K_h$ (due to larger $\sigma$) when the basin width is smaller than $L_A$ but far larger than zero, in which case, the basin size may be too narrow to breed its natural oscillation. Therefore, when the basin size varies from the reasonable range (e.g., from $L_A$ to $L_P$), the fundamental period can be simply seen as the linear function of the basin width.

The amplitude of SSTA can be expressed as

$$I(L) = \max(T) \leq \left| \sum_{n=0}^{\infty} f_n \cos \lambda_n x \right| + \left| \sum_{n=0}^{\infty} \frac{K_h}{\omega_n} g_n \cos \lambda_n x \right|$$

$$\leq \max(T_0) + \frac{K_h}{\omega_1} \max(h_0) \quad , \quad (14)$$

$$= \max(T_0) + P_1(L) \frac{K_h(L)}{2\pi} \max(h_0)$$

where $\max(T_0)$ and $\max(h_0)$ are the maximum values of the initial SSTA and TDA. Eq. (14) suggests that the amplitude of SSTA variation is determined by the amplitudes of the initial values and the basin size. The latter term shows the amplification effect of the air-sea coupling strength on the initial SSTA. Since both fundamental period and air-sea coupling strength increase with the basin size, the amplitude of SSTA will also increase with the basin size (Fig. 3b). Their near linear relation suggests the dominant role of the fundamental period and hence basin width. This can provide an interpretation for the strong air-sea interaction in the Pacific while relatively weak in the Atlantic and Indian Ocean.

**5 Conclusions and discussion**

This paper investigates tropical air-sea interaction through a simple model that is based on the fundamental dynamics of the atmosphere and ocean. The model concludes tropical air-sea interaction to a standard wave equation that can be analytically solved by applying cosine series expansion. The solution suggests that tropical air-sea interaction can be decomposed into a series of eigenmodes, each of which oscillates with its natural frequency. For tropical Pacific, the first and the second spatial modes are necessary to feature the spatiotemporal diversity of El Niño events. For tropical Atlantic and Indian Ocean, the first spatial mode is necessary to depict the Atlantic Niño and the Indian Ocean Dipole. The constant mode is also necessary to represent the consistent variations throughout the tropical Atlantic and Indian Ocean basin. Furthermore, the first mode in the tropical Pacific oscillates with the fundamental period of around 4 years, consistent with the significant oscillation period of the QQ mode. The second mode oscillates with a natural period of around 2 years, half of the fundamental period, also consistent with the QB mode. The first mode in the tropical Atlantic and Indian Ocean oscillates with the fundamental period of around 2 years, both consistent with the existing observations. The near linear dependence of the period and the amplitude on the basin size is also discussed. Providing unchanged air-sea coupling strength, the fundamental period and the amplitude is basically proportional to the basin size. This may explain the longest period and strongest amplitude of the air-sea interaction in Pacific.

Tropical air-sea interaction is the most dominant signal of the interannual climate variability. Compared with large amounts of previous investigations that deal with the spatial and temporal variations separately, this analytical model provides a systematic viewpoint for complex spatiotemporal variations in different tropical oceans. It also provides an explicitly linkage between the amplitude and period and the basin size. To capture the essence of the tropical air-sea interaction, the model ignores the effects of the zonal advection, the net surface heat flux forcing, and the nonlinearity. It also ignores the interactions among the tropical oceans. Our unified and comprehensive understanding for the tropical air-sea interaction will become more and more in-depth as the model is continuously equipped with these mechanisms.

**Reference**

An, S., Park, J., Kim, B., Timmermann, A., & Jin, F. (2012). Impacts of ocean gateway and basin width on Tertiary tropical climate variability in a prototype model, *THEOR APPL CLIMATOL*, 107(1-2), 155-164. DOI: 10.1007/s00704-011-0469-x.

Battisti, D.S., & Hirst, A.C. (1989). Interannual Variability in a Tropical Atmosphere–Ocean Model: Influence of the Basic State, Ocean Geometry and Nonlinearity, *Journal of Atmospheric Sciences*, 46(12), 1687-1712. DOI: 10.1175/1520-0469(1989)046<1687:IVIATA>2.0.CO;2.

Bejarano, L., & Jin, F. (2008). Coexistence of Equatorial Coupled Modes of ENSO, *J CLIMATE*, 21(12), 3051-3067. DOI: 10.1175/2007JCLI1679.1.

Bjerknes, J. (1969). Atmospheric Teleconnections from the Equatorial Pacific, *MON WEATHER REV*, 97(3), 163-172. DOI: 10.1175/1520-0493(1969)097<0163:ATFTEP>2.3.CO;2.

Cai, W., Wu, L., Lengaigne, M., Li, T., McGregor, S., Kug, J., et al. Chang, P. (2019). Pantropical Climate Interactions, *SCIENCE*, 363(6430). DOI: 10.1126/science.aav4236.

Cane, M.A., & Zebiak, S.E. (1985). A Theory for El Niño and the Southern Oscillation, *SCIENCE*, 228(4703), 1085-1087. DOI: 10.1126/science.228.4703.1085.

Carton, J.A., Chepurin, G.A., & Chen, L. (2018). SODA3: A New Ocean Climate Reanalysis, *J CLIMATE*, 31(17), 6967-6983. DOI: 10.1175/JCLI-D-18-0149.1.

Deser, C., Alexander, M.A., Xie, S., & Phillips, A.S. (2010). Sea Surface Temperature Variability: Patterns and Mechanisms, *ANNU REV MAR SCI*, 2(1), 115-143. DOI: 10.1146/annurev-marine-120408-151453.

Gualdi, S., Guilyardi, E., Navarra, A., Masina, S., & Delecluse, P. (2003). The Interannual Variability in the Tropical Indian Ocean as Simulated by a CGCM, *CLIM DYNAM*, 20(6), 567-582. DOI: 10.1007/s00382-002-0295-z.

Hirst, A.C. (1986). Unstable and Damped Equatorial Modes in Simple Coupled Ocean-Atmosphere Models, *Journal of Atmospheric Sciences*, 43(6), 606-632. DOI: 10.1175/1520-0469(1986)043<0606:UADEMI>2.0.CO;2.

Iizuka, S., Matsuura, T., & Yamagata, T. (2000). The Indian Ocean SST Dipole

Simulated in a Coupled General Circulation Model, *GEOPHYS RES LETT*, 27(20), 3369-3372. DOI: https://doi.org/10.1029/2000GL011484.

Jansen, M.F., Dommenget, D., & Keenlyside, N. (2009). Tropical Atmosphere–Ocean Interactions in a Conceptual Framework, *J CLIMATE*, 22(3), 550-567. DOI: 10.1175/2008JCLI2243.1.

Jin, F., & Neelin, J.D. (1993). Modes of Interannual Tropical Ocean–Atmosphere Interaction—a Unified View. Part I: Numerical Results, *Journal of Atmospheric Sciences*, 50(21), 3477-3503. DOI: 10.1175/1520-0469(1993)050<3477:MOITOI>2.0.CO;2.

Jin, F., Chen, H., Zhao, S., Hayashi, M., Karamperidou, C., Stuecker, M.F., et al. Geng, L. (2020), Simple ENSO Models, pp. 119-151, Eds., Michael J. McPhaden, Agus Santoso, Wenju Cai, El Niño Southern Oscillation in a Changing Climate.

Jin, F. (2022). Toward Understanding El Niño Southern-Oscillation's Spatiotemporal Pattern Diversity, *FRONT EARTH SC-SWITZ*, 10.

Jin, F.F. (1996). Tropical Ocean-Atmosphere Interaction, the Pacific Cold Tongue, and the El Niño-Southern Oscillation, *SCIENCE*, 274(5284), 76-78. DOI: 10.1126/science.274.5284.76.

Jin, F.F. (1997). An Equatorial Ocean Recharge Paradigm for ENSO. Part I: Conceptual Model, *Journal of Atmospheric Sciences*, 54(7), 811-829. DOI: 10.1175/1520-0469(1997)054<0811:AEORPF>2.0.CO;2.

Kajtar, J.B., Santoso, A., England, M.H., & Cai, W. (2017). Tropical Climate Variability: Interactions Across the Pacific, Indian, and Atlantic Oceans, *CLIM DYNAM*, 48(7-8), 2173-2190. DOI: 10.1007/s00382-016-3199-z.

Kang, I., & An, S. (1998). Kelvin and Rossby Wave Contributions to the SST Oscillation of ENSO, *J CLIMATE*, 11(9), 2461-2469. DOI: 10.1175/1520-0442(1998)011<2461:KARWCT>2.0.CO;2.

Kang, I., An, S., & Jin, F. (2001). A Systematic Approximation of the SST Anomaly Equation for ENSO, *Journal of the Meteorological Society of Japan. Ser. II*, 79(1), 1-10. DOI: 10.2151/jmsj.79.1.

Keenlyside, N.S., & Latif, M. (2007). Understanding Equatorial Atlantic Interannual

Variability, *J CLIMATE*, 20(1), 131-142. DOI: https://doi.org/10.1175/JCLI3992.1.

Latif, M., & Groetzner, A. (2000). The equatorial Atlantic oscillation and its response to ENSO, *CLIM DYNAM*, 16(2-3), 213-218. DOI: 10.1007/s003820050014.

McPhaden, M.J., Zebiak, S.E., & Glantz, M.H. (2006). ENSO as an Integrating Concept in Earth Science, *SCIENCE*, 314(5806), 1740-1745. DOI: 10.1126/science.1132588.

McPhaden, M.J., & Nagura, M. (2014). Indian Ocean Dipole Interpreted in Terms of Recharge Oscillator Theory, *CLIM DYNAM*, 42(5-6), 1569-1586. DOI: 10.1007/s00382-013-1765-1.

Neelin, J.D., Latif, M., Allaart, M.A.F., Cane, M.A., Cubasch, U., Gates, W.L., et al. Zebiak, S.E. (1992). Tropical Air-Sea Interaction in General Circulation Models, *CLIM DYNAM*, 7(2), 73-104. DOI: 10.1007/BF00209610.

Neelin, J.D., & Jin, F. (1993). Modes of Interannual Tropical Ocean–Atmosphere Interaction—a Unified View. Part II: Analytical Results in the Weak-Coupling Limit, *Journal of Atmospheric Sciences*, 50(21), 3504-3522. DOI: 10.1175/1520-0469(1993)050<3504:MOITOI>2.0.CO;2.

Ng, B., Cai, W., & Walsh, K. (2014). The Role of the SST-thermocline Relationship in Indian Ocean Dipole Skewness and Its Response to Global Warming, *SCI REP-UK*, 4(1). DOI: 10.1038/srep06034.

Prigent, A., Lübbecke, J.F., Bayr, T., Latif, M., & Wengel, C. (2020). Weakened SST Variability in the Tropical Atlantic Ocean Since 2000, *CLIM DYNAM*, 54(5-6), 2731-2744. DOI: 10.1007/s00382-020-05138-0.

Rao, S.A., Behera, S.K., Masumoto, Y., & Yamagata, T. (2002). Interannual Subsurface Variability in the Tropical Indian Ocean with a Special Emphasis on the Indian Ocean Dipole, *Deep Sea Research Part II: Topical Studies in Oceanography*, 49(7), 1549-1572. DOI: https://doi.org/10.1016/S0967-0645(01)00158-8.

Ren, H., & Jin, F. (2013). Recharge Oscillator Mechanisms in Two Types of ENSO, *J CLIMATE*, 26(17), 6506-6523. DOI: 10.1175/JCLI-D-12-00601.1.

Saji, N.H., Goswami, B.N., Vinayachandran, P.N., & Yamagata, T. (1999). A Dipole Mode in the Tropical Indian Ocean, *NATURE*, 401(6751), 360-363. DOI:

10.1038/43854.

Suarez, M.J., & Schopf, P.S. (1988). A Delayed Action Oscillator for ENSO, *Journal of Atmospheric Sciences*, 45(21), 3283-3287. DOI: 10.1175/1520-0469(1988)045<3283:ADAOFE>2.0.CO;2.

Timmermann, A., An, S., Kug, J., Jin, F., Cai, W., Capotondi, A., et al. Zhang, X. (2018). El Niño–Southern Oscillation Complexity, *NATURE*, 559(7715), 535-545. DOI: 10.1038/s41586-018-0252-6.

Vallès Casanova, I., Lee, S.K., Foltz, G.R., & Pelegrí, J.L. (2020). On the Spatiotemporal Diversity of Atlantic Niño and Associated Rainfall Variability Over West Africa and South America, *GEOPHYS RES LETT*, 47(8). DOI: 10.1029/2020GL087108.

Wakata, Y., & Sarachik, E.S. (1991). Unstable Coupled Atmosphere–Ocean Basin Modes in the Presence of a Spatially Varying Basic State, *Journal of Atmospheric Sciences*, 48(18), 2060-2077. DOI: https://doi.org/10.1175/1520-0469(1991)048<2060:UCABMI>2.0.CO;2.

Weller, E., Cai, W., Du, Y., & Min, S. (2014). Differentiating Flavors of the Indian Ocean Dipole Using Dominant Modes in Tropical Indian Ocean Rainfall, *GEOPHYS RES LETT*, 41(24), 8978-8986. DOI: https://doi.org/10.1002/2014GL062459.

Wyrtki, K. (1985). Water Displacements in the Pacific and the Genesis of El Nino Cycles, *Journal of Geophysical Research: Oceans*, 90(C4), 7129-7132. DOI: https://doi.org/10.1029/JC090iC04p07129.

Xie, R., & Jin, F. (2018). Two Leading ENSO Modes and El Niño Types in the Zebiak–Cane Model, *J CLIMATE*, 31(5), 1943-1962. DOI: 10.1175/JCLI-D-17-0469.1.

Zebiak, S.E., & Cane, M.A. (1987). A Model El Niño–Southern Oscillation, *MON WEATHER REV*, 115(10), 2262-2278. DOI: 10.1175/1520-0493(1987)115<2262:AMENO>2.0.CO;2.

Zebiak, S.E. (1993). Air–Sea Interaction in the Equatorial Atlantic Region, *J CLIMATE*, 6(8), 1567-1586. DOI: https://doi.org/10.1175/1520-0442(1993)006<1567:AIITEA>2.0.CO;2.

Zhang, L., & Han, W. (2021). Indian Ocean Dipole Leads to Atlantic Niño, *NAT COMMUN*, 12(1). DOI: 10.1038/s41467-021-26223-w.

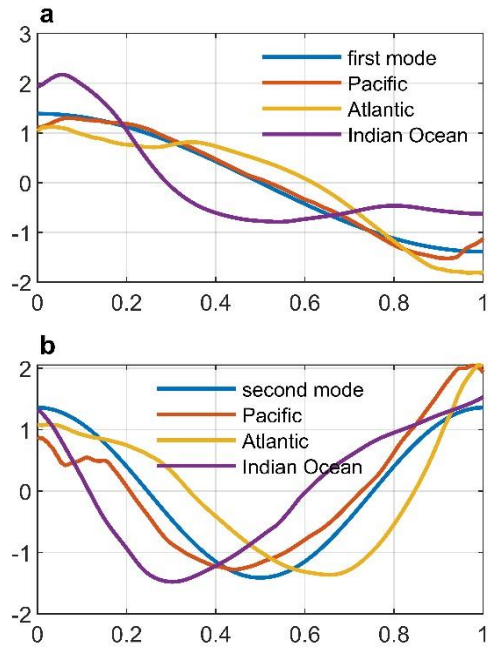

**Figure 1.** First two leading empirical orthogonal function (EOF) patterns and first two cosine expansion modes of the tropical Pacific, Atlantic, and Indian Ocean SSTA variations. The panels show their highly associated normalized patterns in a dimensionless basin size. The EOF analysis is conducted on the monthly tropical (5°S-5°N mean) SSTA evolution of the tropical Pacific (130°E-85°W), the tropical Atlantic (39°W-5°W), and the tropical Indian Ocean (45°E-95°E) in the simple ocean data assimilation (SODA) version 3.4.2 that spans from Jan 1980 to Dec 2020. SSTAs in each month are smoothed by using a 5-grid running mean. (a) The similarity among the first leading EOF mode of the tropical Pacific SSTA, the second leading EOF mode of the tropical Atlantic and Indian Ocean SSTA, and their first cosine expansion mode. (b) The similarity among the second leading EOF mode of the tropical Pacific SSTA, the third leading EOF mode of the tropical Atlantic, Indian Ocean SSTA, and their second cosine expansion mode.

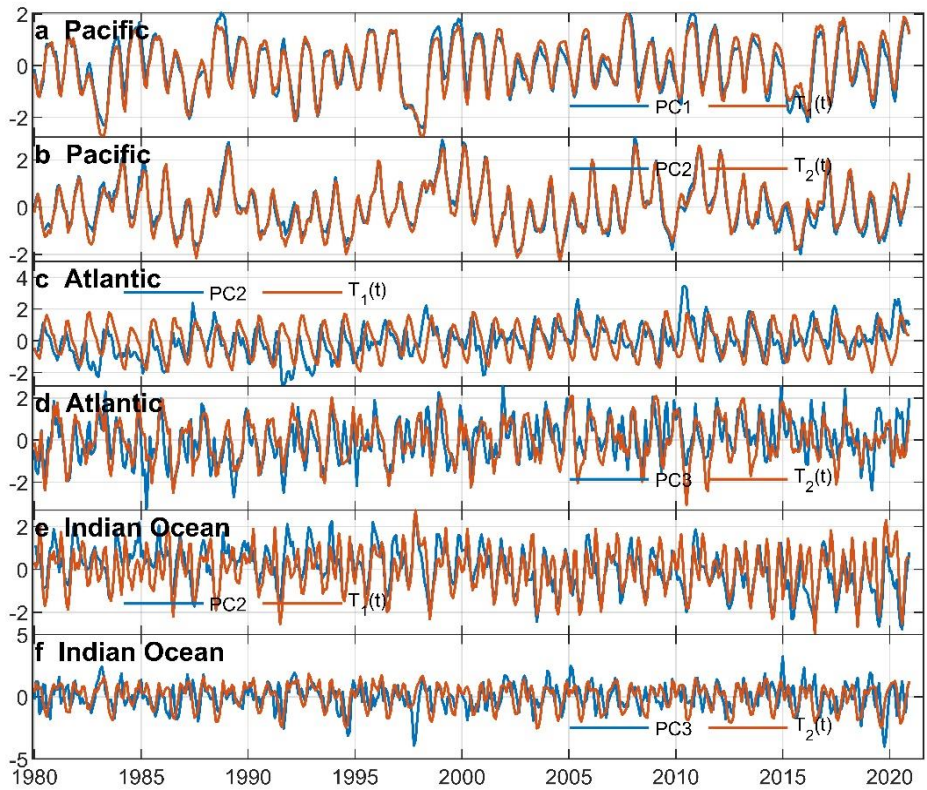

**Figure 2.** Time series of the EOF modes and the cosine expansion coefficients. The panels show that the principal component (PC) time series are highly associated with the corresponding cosine expansion coefficients. (a and b) The first and the second PC time series and the first two cosine expansion coefficients of the tropical Pacific SSTA. (c and d) The second and the third PC time series and the first two cosine expansion coefficients of the tropical Atlantic SSTA. (e and f) Same as (c and d), but for the tropical Indian Ocean SSTA.

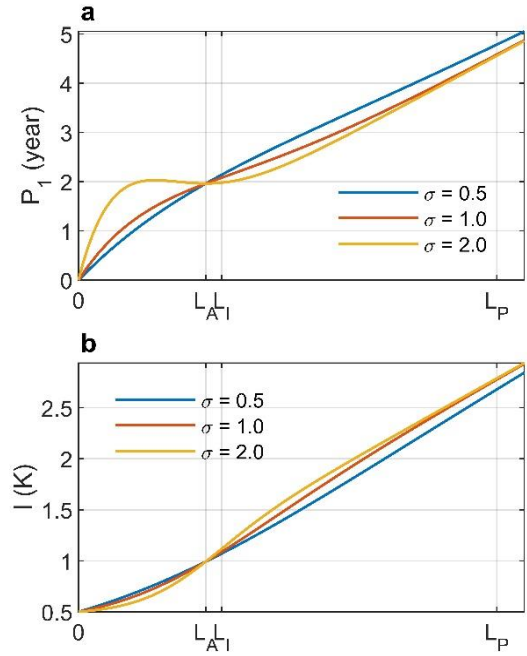

**Figure 3.** The relations between the fundamental period, the amplitude and the basin width. The panels show a near linear relations between the fundamental period, the amplitude and the basin width when the basin width increases from the Atlantic basin width ($L_A$) to the Indian Ocean basin width ($L_I$) and to the Pacific basin width ($L_P$). (a) The fundamental period ($P_1$) as a function of the basin width. (b) The amplitude ($I$) as a function of the basin width. The maximum values of the initial SSTA and TDA are set to be 0.5K and 10m.